# Studying Brazil-Nut Effect History Line using Disk-Formed Objects, Scanner, and Web Browser

Sparisoma Viridi[a,1], Siti Nurul Khotimah[a,2], Novitrian[a,3], Widayani[a,4],
Luman Haris[a,5], Dimas Praja Purwa Aji[a,6]

[a]Nuclear Physics and Biophysics Research Division, Institut Teknologi Bandung, Bandung 40132, Indonesia
[1]viridi@cphys.fi.itb.ac.id, [2]nurul@fi.itb.ac.id, [3]novit@fi.itb.ac.id, [4]widayani@fi.itb.ac.id,
[5]ignlumen@gmail.com, [6]praja_dimas@students.itb.ac.id

**Abstract** - Grains configuration snapshots of Brazil-nut effect (BNE) in two-dimension are physically modeled using disk-formed objects, e.g., buttons and magnetic pin. These BNE configurations are artificially designed to mimic the real ones observed in experiments. A computer scanner is used to capture the configurations. Obtained images are then digitized using web browser running a HTML equipped with a JavaScript code, which is built mainly only for this work. From digitization process all grains positions (granular bed and intruder) are obtained, which is later analyzed using the simplest model, i.e., potential energy. Since the minimum energy principle (MEP) suggests that a closed system should go to its state with minimum internal energy, our BNE system must also obey it. Evolution of only the intruder seems to violate MEP but not for the whole system. Grains compaction plays important role, so that the system can achieve its configuration with minimum potential energy.

## 1. Introduction

Brazil-nut effect (BNE) is phenomenon in granular materials where an intruder (larger grain) can rise in granular bed (smaller grains), while the whole system are introduced to vibration [1]. It can happen in two- [2] and three-dimension [3]. While three-dimension systems are rather difficult to be observed, which require particular observation techniques such as induction coil [4] or visual-mechanical tracer [5], two-dimension system promises better observation visually, where grains configurations can be further recorded and the intruder can also be traced automatically, e.g., using OpenCV application [6].

There is no real BNE phenomenon observed in this work. A grains configuration is designed artificially and grains are positioned based on observed experiments [6-7] and reported research [2-3]. Therefore, the term configuration step $s$ is used instead of real time $t$ in identifying configurations order. Grains configurations are recorder using computer scanner, office equipment which are not so common in use for educational or research purposes. Scanner has been used for simulating roots [8], automatic counting of chemically etched tracks [9], recording color of leaves related to chlorophyll content [10], or as office equipment supporting education process [11-12].

## 2. Experiment

A two-dimension grains configuration is designed using cloths buttons as granular bed and a magnetic pin as intruder, where all of these objects have disk shape and flat surface. These grains (buttons and pin) are put on the scanning area of a computer scanner and are limited using two steel rulers, which act like a granular container. Figure 1 shows the required materials, tools, and apparatus.

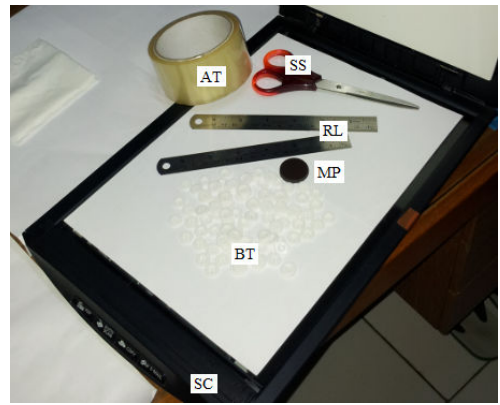

Fig. 1   Required materials, tools, and apparatus: scissor (SS), adhesive tape (AT), rulers (RL), magnetic pin as intruder (MP), cloths buttons as granular bed (BT), and computer scanner (SC).

The two rulers are fixed to the scanner using adhesive tape, so that the granular container has rigid and stable boundaries when the grains are rearranged every time new grains configuration is produced. Color of granular bed and intruder must be different, i.e., making the digitizing process easier. In this work their color are white and black as given in Fig. 1.

Canon CanoScan LiDE 110 is the computer scanner. It has USB connector, A4 size scan area, 3-color (RGB) LED, and about 39.1 s average scan time for each image. In recording the grains configurations 300 dpi resolution and color mode are chosen with output format is multiple pages PDF, which is later converted and cropped into sequence of PNG images using Convert, a Linux application, with options

```
-density 144x144 -resize 1190x1683 -crop 514x1200+0+702
```

Digitizing the image to determine position of all grains (bed particles and intruder) is the next step. A web browser running JavaScript code enhanced HTML is used, where the code is tailored to fit the purpose of this work. Figure 2 shows a screen snapshot of the digitizing application. User can choose an image, digitize it by clicking center of the grains, copy the coordinates into a text file, and save the results. Digitized coordinates are represented in pixels instead of SI unit.

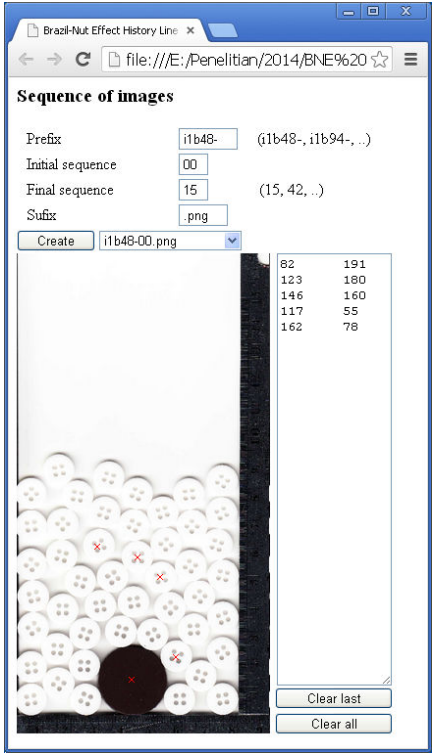

Fig. 2  A internet browser with JavaScript code enhanced HTML is used as a digitizing application in determining grains position at configuration step $s = 00$ for system with one intruder and 48 bed particles.

## 3. Theory

Potential energy $U$ of each grain is given by

$$U = mgy, \quad (1)$$

where $m$ is grain mass, $g$ is gravitation acceleration, and $y$ is vertical position of a grain. For simplicity $g = 1$ is chosen, since $y$ is also represented in pixels, which turns Eq. (1) into

$$U = my\,(\text{px}), \quad (2)$$

where px represents pixel unit. Equation (2) can still give potential energy qualitatively even the units do not agree to common units.

Another interesting feature is distance of center of two grains $\Delta r_{ij}$ (px)

$$\Delta r_{ij} = \sqrt{(x_i - x_j)^2 + (y_i - y_j)^2}, \quad (3)$$

where $x_i$, $x_j$, $y_i$, and $y_j$ are also presented in px unit. If there are $N$ grains in the system, then there will be $M$ different values of $\Delta r_{ij}$, where

$$M = \frac{1}{2} N(N-1). \quad (4)$$

To distribute those values $O$ classes can be defined where width of each class is

$$\Delta O = \frac{\Delta r_{\max} - \Delta r_{\min}}{O}. \quad (5)$$

If the classes are indexed with $k$ from $0 .. (O - 1)$ then class membership a value of $\Delta r_{ij}$ can be determined through

$$k = \left\lfloor \frac{\Delta r_{ij}}{\Delta O} \right\rfloor, \quad (6)$$

where the square brackets with only lower horizontal bars displayed stand for floor function.

## 4. Results and Discussion

It is observed that bed particle and the intruder have diameters $D$ of 32 px and 70 px, respectively. Their real diameters are about 1.2 and 2.6 mm.

Sixteen images are produced for system shown in Fig. 2, which are labeled with configuration step $s = 00, .., 15$. Some examples of the images are given in Fig. 3.

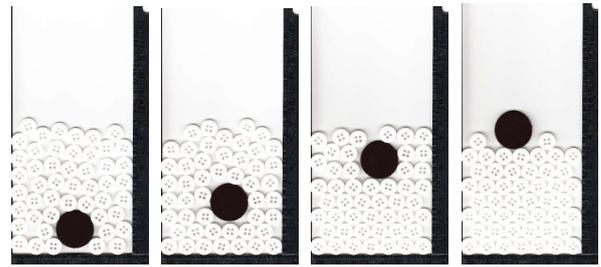

Fig. 3  Grains configurations at step $s = 01, 06, 11,$ and $15$.

At a glance, if only intruder is considered in calculating system potential energy, it seems that system potential energy $U$ is increasing as the configuration step $s$ increasing. Even it is contradicting intuition and naive reasoning, it is not forbidden since energy is being continuously supplied to the system through vibration [13]. Figure 4 is also convincing the increase of $U$ of the intruder.

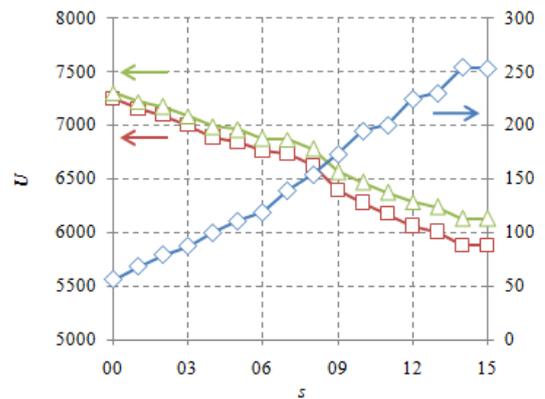

Fig. 4  Potential energy $U$ (in arbitray unit) with grains configuration at step $s$ for intruder ($\Diamond$), granular bed ($\Box$), and whole system ($\triangle$), where $m_i / m_b = 1$.

But if the potential energy for the bed particles is also taken into account, energy the whole system $U$ is decreasing as configuration step $s$ increasing. This tells us that BNE does not violate the minimum energy principle (MEP) [14].

Other interesting property is that the system tries to minimize its packing or every grain, especially bed particles, tries to have as much as possible contact with its neighbors.

This property is similar to contactopy [15], which also increases.

In discussing $\Delta r_{ij}$ only the values related to bed particles are interesting since there is only one intruder in the system. Figure 5 shows distribution of these values for $s = 00$ and 15, where their grains configurations are given in Fig. 6.

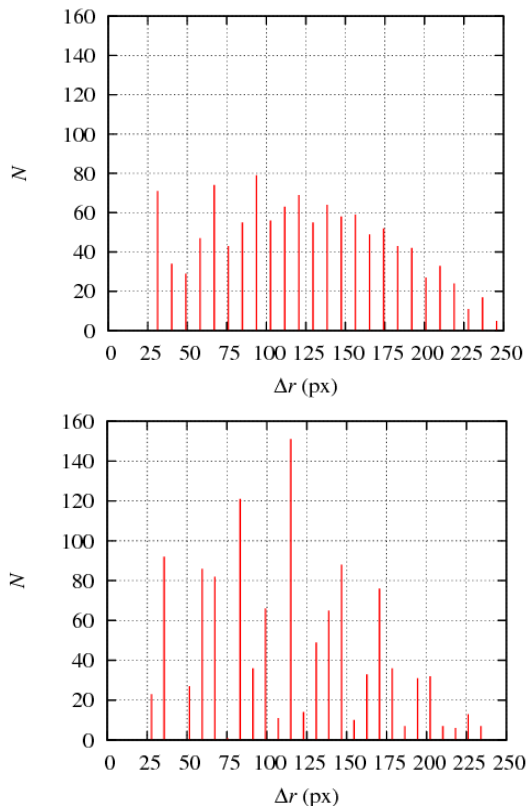

Fig. 5  Distribution of distance of center of two grains $\Delta r_{ij}$ for configuration step $s = 00$ (top) and 15 (bottom) with $O = 30$.

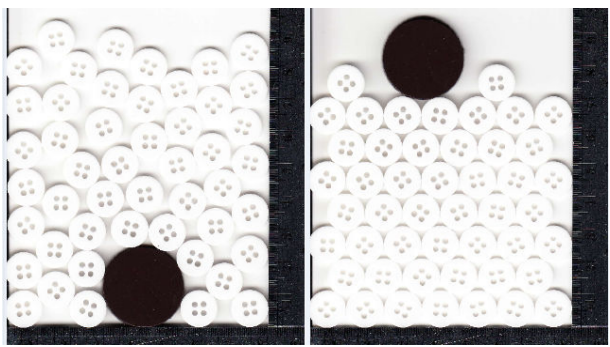

Fig. 6  Grains configuration for configuration step $s$: 00 (left) and 15 (right).

Diameter of bed particle is about 32 px, which should be pronounced in grains configuration with $s = 00$ and also integer multiplication of value 32 px as seen in Fig 5. In grains configuration with $s = 15$, which has a layer of hexagonal close packed (HCP), other value beside 32 px will also be pronounced, which is $\frac{1}{2}\sqrt{3}D$ or about 28 px. These two peaks can be clearly seen as the first two peaks in Fig. 5 (bottom). Existence of other peaks for $s = 15$ can also be found as explained in Fig. 7.

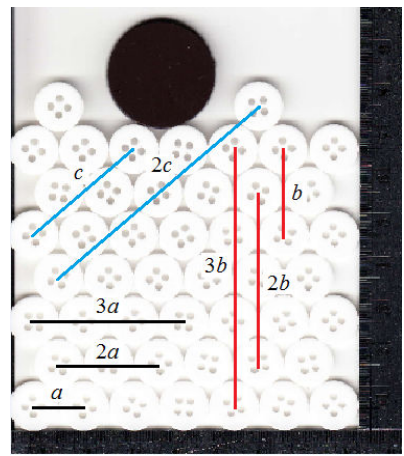

Fig. 7  Various series of $\Delta r_{ij}$: $a$, $2a$, $3a$, .. (black lines), $b$, $2b$, $3b$, .. (red lines), and $c$, $2c$, .. (blue lines) in grains configuration with $s = 15$.

Three series of $\Delta r_{ij}$ can be seen in Fig. 7, which will produce peaks in Fig. 5 (bottom). The most probable peak lies at value about 116 px, which is about $c$ in Fig. 7. Surprisingly this value is more frequent than $a$.

## 5. Conclusion

Artificial grains configurations have been made in order to study potential energy evolution of system mimicked BNE phenomenon. Analyzing potential energy of all grains instead of only the intruder shows us that BNE does not violate MEP. Compaction that tends to HCP can be analyzed using distribution of distance of center of two grains, where shows more pronounce peaks in HCP configuration. The most probable peaks is observed at position about 116 px or about $3.625D$.

## Acknowledgment

This work is supported by Riset Desentralisasi DIKTI (876/AL-J/DIPA/PN/SPK/2014) and RIK-ITB in year 2014 (914/AL-J/DIPA/PN/SPK/2014).